# Is Stubborn Mining Severe in Imperfect GHOST Bitcoin-like Blockchains? Quantitative Analysis

Haoran Zhu, Xiaolin Chang, Jelena Mišić, Vojislav B. Mišić, Lei Han, Zhi Chen

**Abstract**—GHOST, like the longest-chain protocol, is a chain selection protocol and its capability in resisting selfish mining attack has been validated in imperfect blockchains of Bitcoin and its variants (Bitcoin-like). This paper explores an analytical-model-based method to investigate the impact of stubborn mining attack in imperfect GHOST Bitcoin-like blockchains. We first quantify chain dynamics based on Markov chain and then derive the formulas of miner revenue and system throughput. We also propose a new metric, "Hazard Index", which can be used to compare attack severity and also assist attacker in determining whether it is profitable to conduct an attack. The experiment results show that 1) An attacker with more than 30% computing power can get huge profit and extremely downgrade system throughput by launching stubborn mining attack. 2) A rational attacker should not launch stubborn mining attack if it has less than 25% computing power. 3) Stubborn mining attack causes more damage than selfish mining attack under GHOST. Our work provides insight into stubborn mining attack and is helpful in designing countermeasures.

**Index Terms**—Bitcoin-like Blockchain, GHOST, Proof-of-Work, Quantitative Analysis, Stubborn Mining Attack.

## 1 INTRODUCTION

Bitcoin [1] and many of its variants, like Litecoin, Bitcoin Cash and Bitcoin Gold, take Poof-of-Work (PoW) as block generation mechanism, the longest-chain as the chain selection protocol and use miners' block generation reward as their revenue. These blockchains, denoted as **Bitcoin-like blockchains**, have gained the attention of financial sector over the past decade. The value of 1 Bitcoin has increased from $0 in 2009 to $21,418 in January 2022 and has peaked at $66,935 in 2021 [2]. However, Bitcoin-like blockchains are vulnerable to various attacks, affecting blockchain system fairness and security [3]-[5]. Selfish mining attack [6] is a classic one and stubborn mining attack [8] is its variant (detailed in Section 2.1), which increases the hazard to the blockchain [8][9][18]. It is necessary to study the impact of such type threatening attack to prevent financial loss in Bitcoin and Bitcoin-like (**BTC**) blockchain systems.

Most of the existing works only studied selfish mining and stubborn mining attacks in perfect networks [11]-[13][17][18]. However, the actual blockchain network is imperfect [15][16][22]. It takes a period of time for miners to receive newly-found blocks because of the network transmission delay (abbreviated as delay). However, a miner keeps mining until it receives new blocks. As the result, more than one valid block is generated in one block generation interval and then there is a forking occurrence (detailed in Section 2.1). We call this an **unintentional fork**. The fork probability is low in the present Bitcoin (around 0.5%) so that the results in perfect networks are not much different from that in real Bitcoin. Nevertheless, the fork probabilities are higher in other BTC blockchains (1% in Bitcoin Cash and 1.5% in Litecoin [23]). There is still a non-negligible gap between perfect and imperfect networks. Furthermore, the unintentional fork is extremely conducive to attackers and has a huge impact on blockchains [14]. Thus, it is necessary to investigate attacks in imperfect blockchains.

Chain selection mechanism is a basic component in a BTC blockchain to handle fork and the default is the longest-chain protocol, in which only one block is considered when an unintentional fork occurs. Miners choose the longest chain and mine behind the highest block. That is, other blocks are regarded as stale blocks and the computing power is wasted. GHOST [10] is another type of chain selection mechanism applied in Ethereum 2.0 [24] and its variant is used in Ethereum Classic [25]. It considers all published blocks and selects the chain with the most blocks. Fig.1 illustrates these two chain selection protocols, where miners generate blocks on "M6" in the longest-chain protocol and generate blocks on "O7" in GHOST protocol. This is because "M6" is one block higher than "O7" and the chain of "O7" has two more blocks than the chain where "M6" locates. Researchers have investigated selfish mining attack under GHOST in imperfect BTC blockchains [16][21]. **How does stubborn mining attack perform in imperfect BTC blockchains that apply GHOST as the chain selection protocol?**

The above discussions motivate the work of this paper. This paper aims to quantitatively analyze stubborn mining attack in imperfect GHOST BTC blockchains by developing a stochastic model. There exist simulation-based analyses of stubborn mining attack under the longest-chain protocol [20]. However, the simulation-based works are hard to expand to imperfect GHOST blockchains because chain selection protocol affects both chain evolution and miner behaviors. In addition, the authors in [22] made simulations in GHOST Bitcoin blockchain. But they evaluated stubborn mining attack from the perspective of miner revenue and system throughput. We not only assess these metrics but investigate attack severity through our proposed metric "hazard index". Moreover, we compare stubborn mining attack with other classic attacks to illustrate severity of attacks.

---

• *Haoran Zhu and Xiaolin Chang are with the Beijing Key Laboratory of Security and Privacy in Intelligent Transportation, Beijing Jiaotong University, P.R.China.*

Analytical-model-based methods have been proposed for analyzing imperfect scenarios under the longest-chain [14][15] and GHOST [16] but only for selfish mining. Although stubborn mining attack is a variant of selfish mining, miner behaves differently, causing extra system states that do not exist in analytical models of selfish mining. Note that the work of [16] also explored a model of stubborn mining. But their model missed some key states and state transitions so that only describes $T_1$-stubborn (detailed in Section 2.1). They did not give the metrics formulas of stubborn mining attack and their methods for computing metrics were hard to apply to stubborn mining. In addition, researchers have developed analytical models for studying the longest-chain blockchain under stubborn mining attack [17]-[19]. These analytical models are hard, if not impossible, to be used in GHOST.

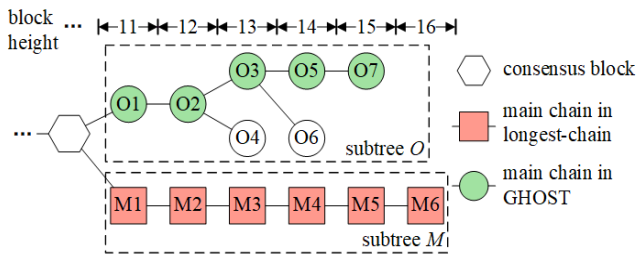

Fig.1 GHOST and the longest-chain protocol

There are at least three major challenges in modeling-based analysis of stubborn mining attack in imperfect GHOST BTC blockchains (detailed in Section 3.1). The major contributions of this paper are summarized as follows.

- We utilize continuous time Markov chain technique to establish an analytical model in order to quantify the evolution and dynamics of imperfect GHOST BTC blockchains. Our model is able to work in Bitcoin and most Bitcoin-like blockchain systems, where a) PoW is used as block generation mechanism, b) miners get revenue by producing valid blocks, c) there exist network-caused forks, and d) GHOST is applied to resolve forks. To the best of our knowledge, we are the first to model and analyze stubborn mining attack in GHOST blockchains with unintentional forks.
- We establish formulas for computing miner revenue and system throughput of all stubborn mining attack strategies, which can be used to evaluate the impact on GHOST BTC blockchain systems. Only consensus blocks are rewarded and only transactions in consensus blocks are considered. However, it depends on all miner behaviors that whether a valid block becomes a consensus block. Attacker behaves differently with different stubborn strategy. It is difficult to calculate the revenues and throughput by existing works. We innovatively derive the formulas of the intermediate variable "subtree selection probability" and enable the metric calculations. To the best of our knowledge, we are the first to derive the formulas for these metrics of all stubborn mining attack strategies in imperfect GHOST scenarios.
- We propose a new metric called "**hazard index**" (HI), namely, extra relative revenue per computing power, to evaluate the threatening severity of attacks. The existing works [11]-[21] used relative revenue (RR) to evaluate both attacker profit and attack severity. However, a higher RR can be caused by larger computing power. That is, RR cannot be used to compare the threat severity of different attack strategies unless the computing power of the attacker is fixed. In addition, RR cannot tell attacker about how profitable it is to conduct an attack. The experiment results in Section 4.3.3 demonstrate the effectiveness of HI.

The rest paper is structured as presenting background and related work in Section 2. Section 3 describes the analytical model of the studied system and metric formulas. Section 4 gives experiment results. Section 5 makes conclusions and discusses future work.

## 2 PRELIMINARIES

### 2.1 Background

**Proof of Work (PoW).** In the PoW block generation mechanism, a **valid block** is generated when there is a solution to a hash-based puzzle. The probability that a miner finds a valid block is proportional to its computing power [6][8]. The time between two adjacent blocks is called **block generation interval**. The average block generation interval is 10 minutes in Bitcoin [1] and 2.5 minutes in Litecoin [7].

**Reward mechanism.** It is an essential function in BTC blockchains. A miner gets a block generation reward after its block reaches the consensus of all miners. A block that is acknowledged by all miners is denoted as **consensus block**. The miner of a consensus block gets one unit of reward. The chain from the **genesis block** (the first block of the blockchain) to the newest consensus block is called the **main chain** and all blocks on the main chain are named **main blocks**. **Height** denotes the number of blocks from the block to the genesis block and the **distance** is the difference in the block height.

We use Fig.1 to illustrate the above concepts. There are already 10 consensus blocks, then the height of "M1" and "O1" is 11. The distance between "M1" and "M3" is 2. Similarly, the distance between "M1" and "O7" is 4. The **length** of a chain is the maximum block height in the chain. In blockchains adopting the longest-chain protocol, the red chain is the longest so that it is the main chain. The **weight** of a block is the number of blocks linked behind that block. A block is a leaf block if the block weight is 0. In Fig.1, the weight of "M2" is 4 and the weight of "O2" is 5. "O4", "O6", "O7" and "M6" are leaf blocks. But honest miners only mine on "M6" in the longest-chain or "O7" in GHOST. The

**leaf block** specifically refers to the blocks that HPs may mine on unless special note. In GHOST, all valid blocks form an acyclic graph, forming a tree structure. Valid blocks behind the consensus block also form several **subtree**s (such as subtree O containing "O1" ~ "O7"). The weight of the root of a subtree is also the weight of the subtree. Subtree O is heavier than subtree M so the root of O ("O1"), and so on the green chain is the main chain. We call the subtree that attackers mine on as **private subtree** and other subtrees are **public subtree**s.

**Mining Pools.** It is very difficult for a single miner to find a PoW solution in blockchains with extremely large total computing power [26]. Miners get nothing until they produce consensus blocks. Thus, miners work in the form of mining pools to gather their computing power. There is a manager and many in-pool miners in a pool. The manager determines the block on which in-pool miners mine and publishes valid blocks produced by in-pool miners. An in-pool miner cannot mine on other blocks otherwise it will be detected as dishonesty and removed from the pool. The revenue is shared equally among all in-pool miners. In other words, if one in-pool miner generates a consensus block, all in-pool miners can get revenue and the total revenue of all in-pool miners is one reward unit.

**Stubborn mining attack.** It is a variant of selfish mining attack proposed in [8]. [22] gives the strategies in imperfect GHOST blockchains. There are 3 basic strategies, Lead (*L*), equal-Fork (*F*), and Trail (*T*), and 4 hybrid strategies, Lead-Fork (*LF*), Lead-Trail (*LT*), Fork-Trail (*FT*), and Lead-Fork-Trail (*LFT*). Attacker can launch stubborn mining attack by adopting any one of the strategies. Note that *T* strategy is a family of strategies. When the private subtree falls behind the public subtree *j* block(s), MP still tries to generate blocks on the private subtree hoping to catch up with the public subtree, denoted as $T_j$ strategy. We only study $T_1$ strategy since it is the dominant strategy in *T* family [8].

Let $\Delta$ denote the weight difference between the private subtree and the heaviest public subtree. $\Delta = 0$ denotes that private and public subtree have same weights. If $\Delta = 0$, the weights of subtrees can be 0 (miners mine on consensus block) or positive (both subtrees exist and denote as 0'). The following are about the differences of the three basic strategies to selfish mining attack.

• *L* strategy: The attacker only publishes the first private block if $\Delta = 2$ and honest miners produce a block.

• *F* strategy: If $\Delta = 0'$ and the attacker generates the next block, it does not publish the block.

• *T* strategy: If $\Delta = 0'$ and honest miners create blocks in public subtree, the attacker still mines on the private subtree. If $\Delta = -1$ and the attacker generates a block, the weights of the two chains are equal. However, all honest miners have been mining on the public subtree, and we use $\Delta = 0''$ to denote this case.

The behaviors of the attacker in selfish mining attack (denoted as "*S*" in the following) and 3 basic strategies of stubborn mining attack are summarized in TABLE 1, where the *italics* denote different behaviors. The behaviors of hybrid strategies are combination of the different behaviors of basic strategies.

## 2.2 Related Works

*Selfish mining attack*. Since selfish mining attack proposed in [6], numerous studies have been working on the impact on blockchain systems. Davidson et al. [11] made simulation-based investigation of the selfish mining profitability under different difficulty adjustment algorithms. Lee et al. [12] studied selfish mining in terms of pools' revenue by a state machine. Ai et al. [13] proposed an optimized selfish mining attack ESM and evaluated relative revenue through simulation and state machine. The results of [11]-[13] are given in perfect networks that do not have unintentional forks.

Authors in [14][15] discussed selfish mining in imperfect networks. They investigated the relative revenue of pools and system throughput in Bitcoin by Markov-chain-based models. All the above researches study the blockchains using the longest-chain protocol. Yang et al. [16] made a Markov-chain-model of selfish mining in an imperfect GHOST blockchain. They made a comparison of GHOST and the longest-chain protocol. This is the only work that comes close to ours. They also explored a method to model stubborn mining attack. But their model misses some key system states and lacks many state transitions so that it can only capture *T* strategy. They did not derive the metrics formulas for stubborn mining attack. Moreover, their method of calculating metrics for selfish mining is hard to apply in stubborn mining attack.

*Stubborn mining attack in perfect networks*. As a variant of selfish mining, stubborn mining attack also attracts many researchers. Grunspan et al. [17] studied *L* and *F* strategies in a perfect network by applying a state machine. The authors in [18] evaluated stubborn mining attack from the perspective of the relative revenue of the attacker, stale block ratio, throughput, and security metric. They built a Markov chain and derived the metrics formulas. Xia et al. [21] did simulations to study stubborn mining attack in GHOST and investigated the performance of selfish mining and *LFT* strategy in a perfect network. These works are based on the perfect network assumption, which may cause a gap from reality.

*Stubborn mining attack in imperfect networks* Wang et al. [19] modeled the selfish mining attack, *T* and *FT* strategies in imperfect PoW Ethereum. They focused on the uncle reward (a unique reward mechanism in Ethereum) and omitted some state transitions that were irrelevant to the uncle reward. Their definition of "imperfect" is different from ours. The miners do make forks but they choose blocks based on timestamp. A block in one generation interval cannot be the main block unless it has the earliest timestamp. In other words, their work is based on an implicit assumption that most miners have perfect networks and few miners have connection errors, which makes the whole network imperfect. We consider a global imperfect network where all miners create forks and every block has a probability to be a main block. Liu et al. [20] investigated

TABLE 1
SELFISH MINING AND BASIC STRATEGIES OF STUBBORN MINING ATTACK IN GHOST BTC BLOCKCHAINS

| State | Attacker finds a block | | | | Honest miners find $i$ blocks in public subtree ($i>0$) | | | | | Honest miners find block(s) in private subtree | | |
|---|---|---|---|---|---|---|---|---|---|---|---|---|
| | $\Delta \geq 0$ | $\Delta = 0'$ | $\Delta = 0''$ | $\Delta = -1$ | $\Delta \geq i+2$ | $\Delta = i+1$ | $\Delta = i$ | $\Delta = i-1$ | $\Delta \leq i-2$ | $\Delta > 0$ and $\Delta \neq 2$ | $\Delta = 2$ | $\Delta = 0$ |
| S | publish | | | | | publish all | | mine pb | | | publish all | |
| L | publish | hold | - | - | publish $i$ | *publish i* | publish all | mine pb | mine pb | publish 1 | *publish 1* | mine on the new block in pri |
| F | hold | | | | | publish all | | | | | publish all | |
| T | publish | *publish* | *publish* | | | publish all | | *mine pri* | | | publish all | |

* "pb" denotes the public subtree, "pri" denotes the private subtree.

TABLE 2
DEFINITION OF NOTATIONS

| Notation | Definition |
|---|---|
| $\Delta$ | The weight difference between the private subtree and the heaviest public subtree. |
| $Hs$ | The number of the subtree(s) that HPs generate. |
| $N$ | The number of leaf block(s) where HPs can mine. |
| $\theta$ | Probability of unintentional fork. |
| $\alpha, \beta$ | The total computing power of MP and HPs, respectively. They also represent the rates of MP and HPs generating block(s). |
| $\beta_1, \beta_2$ | Rates of HPs generating one block and two blocks in a block generation interval, respectively. $\beta_2 = \beta \cdot \theta$, $\beta = \beta_1 + 2\beta_2$. |
| $P_{\beta_1}$ | Probability of HPs generating one block. $P_{\beta_1} = \beta(1-\theta)$. |
| $\gamma_n$ | Probability that HPs generate one block on private subtree when there are $n$ leaf blocks. |
| $g_N^M$ | Probability that HPs generate two blocks on $M^*$ when there are $N$ leaf blocks. |
| $P_A(\Delta, Hs, N)$ | Probability that the first block of private (Attackers') subtree becomes the main block when the system is in state $(\Delta, Hs, N)$. |
| $P_H(\Delta, Hs, N)$ | Probability that the first block of public (Honest miners') subtree becomes the main block when the system is in state $(\Delta, Hs, N)$. |
| $\pi(\Delta, Hs, N)$ | Steady-state probability of state $(\Delta, Hs, N)$. |
| $E_M, E_H$ | Revenue expectations of MP and HPs, respectively. |
| $RR_M, RR_H$ | Relative revenue of MP and HPs, respectively. |
| TPS | Transactions per second. |

* $M \in \{A, AH, H, H1, H2\}$ denotes the fork scenarios. A, AH, H, H1 and H2 denote HPs generate two blocks behind attackers' subtree (private subtree), attackers' and one honest subtree (public subtree), honest subtree(s), one honest subtree, and two honest subtrees, respectively. $g_3^H = g_3^{H1} + g_3^{H2}$

the impact of stubborn mining toward imperfect PoW Ethereum by simulation. All the above works are based on the longest-chain protocol.

The authors in [22] proposed the stubborn mining attack strategies in imperfect GHOST blockchains and evaluated miner profit and system throughput. We not only assess these metrics but also investigate attack severity through hazard index. We make comparison among stubborn mining attack and other classic attacks. However, simulation methods have to restore the reality exactly and have the issue of large rounds of experiments and unstable results.

In summary, the existing works consider stubborn mining attack in the longest-chain protocol, or consider some of stubborn mining strategies in GHOST, or get results only from simulations. Moreover, the existing models are hard, if not impossible, to be used to analyze all 7 stubborn mining strategies in our scenarios. In this paper, we study the impact of all 7 strategies of stubborn mining attacks in imperfect GHSOT BTC blockchains by Markov chain model. We also make a simulator to cross-validate our results.

## 3 SYSTEM DESCRIPTION AND MODEL

This section first describes the system to be studied and the challenges of quantitative analysis in Section 3.1. Then the model and formulas are presented in Sections 3.2 and 3.3, respectively. TABLE 2 shows the notations to be used in the following.

### 3.1 System Description

The PoW is taken as the block generation mechanism and the GHOST is as chain selection protocol in BTC blockchains. The network is imperfect and then unintentional forks can appear. Pools get block generation reward as their revenue.

There are two types of miners, malicious and honest miners. All malicious miners conspire and form a malicious pool (MP). The left miners (namely, honest miners) form several honest pools (HPs). The MP conducts stubborn mining attack and it can quickly detect new blocks from other pools and propagate its block through the assistance of eclipse attack and network sniffing [8][20]. Thus, it is reasonable and feasible to assume that an MP is well-connected and does not make unintentional forks. HPs mine blocks by using the honest mining strategy, detailed as follows.

- Pools mine blocks continuously on the subtree which they think is the heaviest.
- Pools publish blocks as soon as they find them. That is, pools do not withhold any block.
- Pools may create unintentional forks caused by the imperfect network. In other words, at least one block is generated in one block generation interval.
- Pools resolve forks according to GHOST. When there is more than one subtree with the same weight, pools choose the one that contains their own block if exists. Otherwise, they select a subtree randomly.

Computing power is a key property of a pool, which determines the probability of the pool producing blocks in PoW systems. In BTC blockchains, the block generation probability is proportional to the computing power. In fact, the computing power of the largest pools in BTC blockchains is around 17%~30% [26]. A pool with more than half of computing power can conduct double-spending attack and get all revenue, which is unnecessary to study. In addition, the existing works show that small pools (pools with low computing power) conducting stubborn mining attack get no benefits [8][18]. To make our work comprehensive and meaningful, we study the scenarios that the computing power of a pool is between 5% and 45%. The results can show the profit of the stubborn mining attack on large pools and the loss on small pools. There are several assumptions in the system we consider.

- The total computing power in the system is constant, and all miners keep mining blocks.
- There is only one group of attackers in the system. They only conduct stubborn mining attack.
- There is no individual miner in the system. A single miner can be regarded as a small pool composed of one miner, where the in-pool miner is also the manager.
- No more than two blocks can be created simultaneously due to the small probability of the occurrence of a three-chain fork [27]. Thus, there are two new blocks generated by HPs when an unintentional fork occurs.

There are at least the following challenges to system modeling and formula calculating.

**Challenge 1.** The behaviors of MP and HPs are interrelated. MP behaves differently in different stubborn strategies. How to quantify their interaction in one model is the first challenge.

**Challenge 2.** When HPs generate a fork, the two newly generated blocks can locate in different or same subtree. The blocks can also locate in public or private subtrees. These situations combine randomly, leading to different blockchain structures. How to characterize these scenarios is the second challenge.

**Challenge 3.** In the longest-chain protocol, MP launching selfish mining attack is rewarded for sure in some situations (such as $\Delta = 2$ and HPs find blocks). However, these situations are not fully established in GHOST. Stubborn mining attack makes the system fork instead of getting a consensus. Forked system is hard to calculate revenue accurately. How to compute the revenue of stubborn pools in GHOST is a huge challenge.

### 3.2 System Model

This section presents an analytic model of the chain dynamics under stubborn mining attack in a GHOST BTC system. The notations and definitions are summarized in TABLE 2.

We assume the block generation time following exponential distribution as in [13]-[16]. A three-tuple $(\Delta, Hs, N)$ is used to denote a system state:

- $\Delta$ denotes the difference in the weight between the private subtree and the heaviest public subtree. It is a key element that determines the attackers' behavior.
- $Hs$ denotes the number of subtrees that HPs generate. It also represents the number of public subtrees. Due to the feature of "imperfect", HPs can generate more than one public subtree. Pools can mine blocks on any of the subtrees, which means the dispersion of computing power and benefits the attacker. Note that a subtree is defined as the branch behind the consensus block. Thus, $Hs$ could be 0, representing no new blocks generated by HPs after the consensus block.
- $N$ denotes the number of blocks that pools can mine on, including blocks on private subtree, blocks on public subtree and the consensus block.

With the assumptions in Section 3.1, $\Delta \in \{-1, 0", 0, 1, ...\}$, $Hs \in \{0, 1, 2\}$, and $N \in \{1, 2, 3\}$. There are constraints on the tuples listed as follows.

1) MP only produces blocks on private subtree and does not produce unintentional fork. Thus, there is only one private subtree and only one leaf block on it.
2) Since at least one block exists in a subtree, $N \geq Hs$.
3) Two new blocks are generated when an unintentional fork occurs, $N - Hs \leq 2$.
4) In $T$ strategy, MP insists on mining on the private subtree although the weight of the public subtree is greater than that of the private subtree. In this scenario, all HPs mine on the public subtree and cannot see the private subtree. $N = Hs$ although the private subtree has been published.

Fig.2 describes the state-transition diagram, where the details in the black dotted frame in Fig.2 (a) are displayed in Fig.2 (b). Now we present how to calculate the transition rates of Fig.2. The total computing power of the system varies greatly in different BTC blockchains (286.56Eh/s in Bitcoin and 1.49Eh/s in Bitcoin Cash [26]). The block generation rate is also different in BTC blockchains (1/600 in Bitcoin and 1/150 in Litecoin). Without loss of generality, we normalize the block generation rate and the total computing power of the system to 1. That is, one block can be generated per time unit on average. We define the probability of creating an unintentional fork as $\theta$. $\alpha$ and $\beta$ denote the total computing power of MP and HPs, respectively. They also represent the rates of MP and HPs generating block(s). $\alpha + \beta = 1$. Recall that the total block generation rate is 1 and MP does not generate forks. Thus, the rate of pools generating two blocks is $\beta_2 = \beta \cdot \theta$ and the total rate of all pools generating one block is $r_1 = 1 - 2\beta_2$.

There are 7 different strategies in stubborn mining attack so the state transitions are different. In Fig.2, some lines are split in two and one of them is marked with a word. For example, the state (2,2,3) transits to (0,0,1) with a green line and transits to (1,1,2) with a green line with the mark "$L$". This denotes that if the attacker applies $L^*$ strategy (namely, $L$, $LF$ and $LFT$), the state (2,2,3) transits to (1,1,2) at a rate of $\beta_1$. Otherwise, the state (2,2,3) transits to (0,0,1) at the same rate. Similarly, the state (2,1,2) transits to (1,1,2) at a rate of $\beta_1$ if MP applies $L^*$ strategy or it transits to (0,0,1) at the same rate.

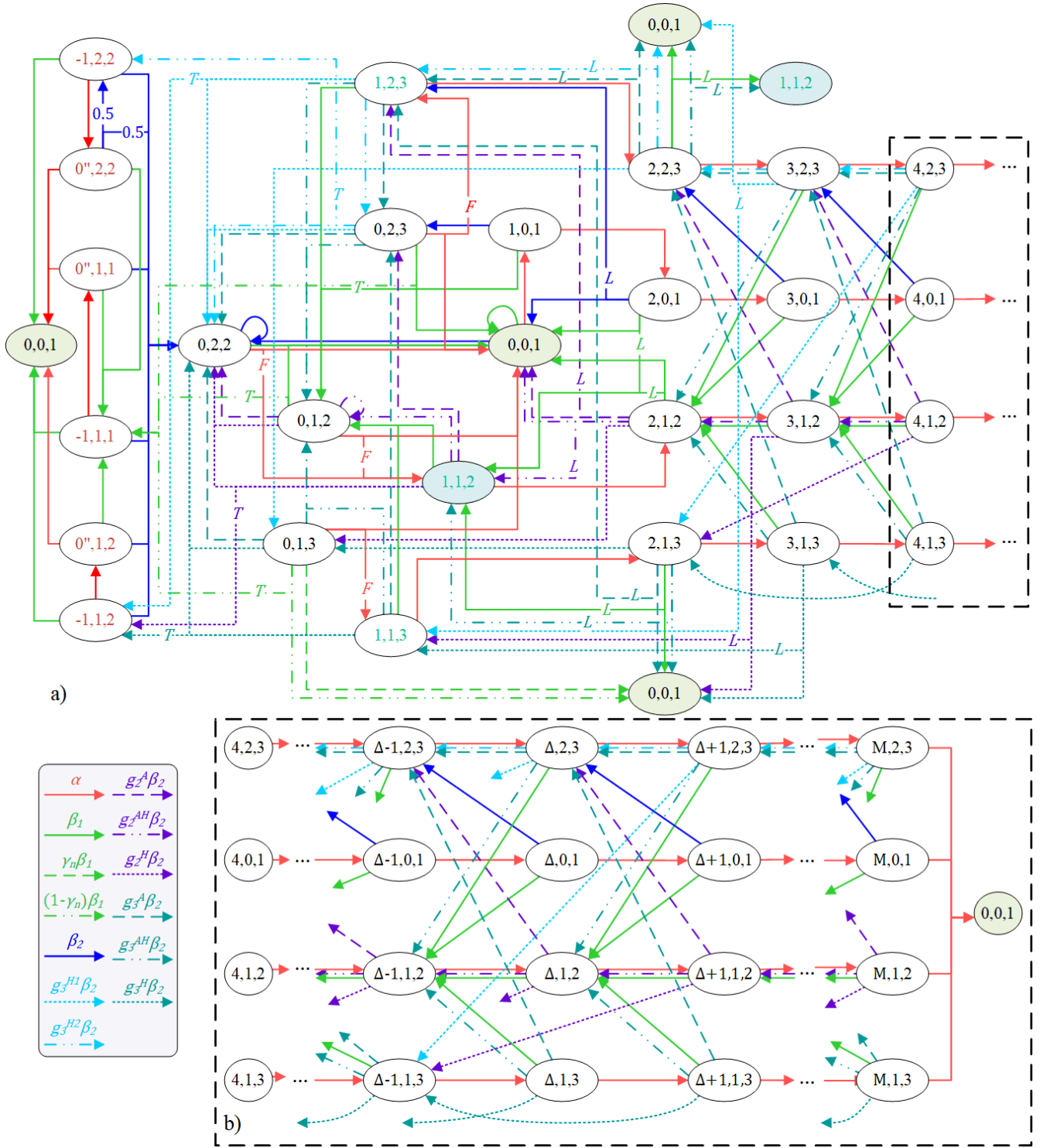

Fig.2 State-transition diagram

### 3.3 Metric Formulas

Let $\pi(\Delta, Hs, N)$ denote the steady-state probability of the state $(\Delta, Hs, N)$. We can get the steady-state probabilities of all states by solving the global balance equations and apply the results to calculate metrics. This section first calculates the revenues of MP and HPs in GHOST BTC blockchains. Then we use this metric to calculate the relative revenue of pools, transactions per second, and hazard index of stubborn mining attack.

#### 3.3.1 Revenue of pools

The revenue of pools is a weighted average of the expected revenue of pools in each system state. The weight is the steady-state probability of each state. We consider all block generation events (denoted as $x$) in each state, including MP generating a block, HPs generating a block, and HPs generating two blocks. The revenue of MP (denoted as $E_M$) and HPs (denoted as $E_H$) can be derived by Eqs. (1)-(2).

$$E_M = \sum_{(\Delta, Hs, N)} \pi(\Delta, Hs, N) \sum_x G(x) \cdot n \cdot P_A(S(x)) \quad (1)$$

$$E_H = \sum_{(\Delta, Hs, N)} \pi(\Delta, Hs, N) \sum_x G(x) \cdot n \cdot P_H(S(x)) \quad (2)$$

where $G(x)$ is the rate of block generation event $x$, $n$ is the number of blocks published by the pool, $P_A$ and $P_H$ are subtree selection probability, and $S(x)$ is the new system state after the event $x$.

We now give the definition and calculation of the subtree selection probability, which is the only unknown variable in Eqs. (1)-(2). When there are both public subtree and private subtree, only one of the subtrees can be the heaviest subtree and the first block of that subtree becomes consensus block. We call the probability that the first block of one subtree becomes the consensus block as subtree selection probability. Let $P_A(\Delta, Hs, N)$ and $P_H(\Delta, Hs, N)$ denote the probabilities that the first block of private (Attackers') and public (HPs') subtree becomes the main block in state $(\Delta, Hs, N)$, respectively. $P_A(\Delta, Hs, N) + P_H(\Delta, Hs, N) = 1$. $Pb(\Delta, Hs, N)$ is defined in Eq. (3) for auxiliary calculations of subtree selection probability. The formulas for calculating subtree selection probability are given in **Proposition 1**.

**Proposition 1.** *The subtree selection probability can be derived by solving Eqs. (3)-(4).*

$$\begin{cases} Pb(0,1,N) = P_{\beta 1}(\gamma_N + (1-\gamma_N)P_A(-1,1,N-1)T) + \beta_2(g_N^A \\ + g_N^{AH} P_A(0,1,2)) \\ Pb(1,1,N) = P_{\beta 1}(\gamma_N + (1-\gamma_N)P_A(0,1,2)) + \beta_2(g_N^A + g_N^{AH} \\ + g_N^H P_A(-1,1,N-1)T) \\ Pb(2,1,N) = P_{\beta 1}(\gamma_N + (1-\gamma_N)((1-L) + P_A(1,1,2)L)) \\ + \beta_2(g_N^A + g_N^{AH} + g_N^H P_A(0,1,3)) \\ Pb(3,1,N) = P_{\beta 1}(\gamma_N + (1-\gamma_N)P_A(2,1,2)) + \beta_2(g_N^A + g_N^{AH} \\ + g_N^H((1-L) + P_A(1,1,3)L)) \\ Pb(\Delta,1,N) = P_{\beta 1}(\gamma_N + (1-\gamma_N)P_A(\Delta-1,1,2)) + \beta_2(g_N^A \\ + g_N^{AH} + g_N^H P_A(\Delta-2,1,3)) \\ Pb(0,2,3) = P_{\beta 1}(\gamma_3 + (1-\gamma_3)P_A(-1,1,1)T) + \beta_2(g_3^A + g_3^{AH} \\ P_A(0,1,2) + g_3^{H2} P_A(-1,2,2)T) \\ Pb(1,2,3) = P_{\beta 1}(\gamma_3 + (1-\gamma_3)P_A(0,1,2)) + \beta_2(g_3^A + g_3^{AH} \\ + g_3^{H2} P_A(0,2,3) + g_3^{H1} P_A(-1,1,2)T) \\ Pb(2,2,3) = P_{\beta 1}(\gamma_3 + (1-\gamma_3)((1-L) + P_A(1,1,2)L)) + \beta_2( \\ g_3^A + g_3^{AH} + g_3^{H2}((1-L) + P_A(1,2,3)L) + g_3^{H1} P_A(0,1,3)) \\ Pb(3,2,3) = P_{\beta 1}(\gamma_3 + (1-\gamma_3)P_A(2,1,2)) + \beta_2(g_3^A + g_3^{AH} \\ + g_3^{H2} P_A(2,2,3) + g_3^{H1}((1-L) + P_A(1,1,3)L)) \\ Pb(\Delta,2,3) = P_{\beta 1}(\gamma_3 + (1-\gamma_3)P_A(\Delta-1,1,2)) \\ + \beta_2(g_3^A + g_3^{AH} + g_3^{H2} P_A(\Delta-1,2,3) + g_3^{H1} P_A(\Delta-2,1,3)) \\ \Delta \geq 4 \end{cases} \quad (3)$$

$$\begin{cases} P_A(0, Hs, N) = Pb(0, Hs, N) + \alpha(1-F) \\ +F \sum_{i \geq 1} \alpha^i \cdot Pb(i, Hs, N) \\ P_A(\Delta, Hs, N) = Pb(\Delta, Hs, N) + \alpha P_A(\Delta+1, Hs, N)(1-F) \\ +F \sum_{i \geq 1} \alpha^i \cdot Pb(i+\Delta, Hs, N) \\ P_A(0'',1,1) = \alpha / (1 - P_{\beta 1}\alpha) \\ P_A(0'',2,2) = \alpha / (1 - (P_{\beta 1} + 1/2\beta_2)\alpha) \\ P_A(0'',1,2) = \alpha / (1 - P_{\beta 1}\alpha) \\ P_A(-1,1,1) = \alpha^2 / (1 - P_{\beta 1}\alpha) \\ P_A(-1,2,2) = \alpha^2 / (1 - (P_{\beta 1} + 1/2\beta_2)\alpha) \\ P_A(-1,1,2) = \alpha^2 / (1 - P_{\beta 1}\alpha) \\ \Delta \geq 0, (Hs, N) \neq (0,1) \end{cases} \quad (4)$$

where $P_{\beta 1}$ is the probability that HPs generate a block and $L, F, T$ are flags of MP's strategy. Their values are 1 if the attacker adopts corresponding strategies, otherwise are 0. For example, if MP adopts *LT* strategy, $L=1$, $F=0$, $T=1$.

### 3.3.2 Relative revenue of pools

The increase in the revenue of MP does not mean a decrease in the revenue of HPs, and vice versa. The total revenue can vary with the attack and fork probability. Thus, we calculate the relative revenue (RR) of pools to measure the proceeds from launching stubborn mining attack. A pool's relative revenue is defined as the proportion of the pool's revenue in all revenues. The relative revenue of HPs ($RR_H$) decreases if the relative revenue of MP ($RR_M$) increases. Eq. (5) can calculate $RR_M$ and $RR_H$.

$$\begin{cases} R_T = E_M + E_H \\ RR_M = E_M / R_T \\ RR_H = E_H / R_T \end{cases} \quad (5)$$

### 3.3.3 Transactions per second

Similar to [14] and [15], we use transactions per second (TPS) to evaluate the system throughput. Only the transactions in main blocks can be accounted for in BTC blockchains. The number of transactions in a block is almost constant [26], and thus the TPS can be evaluated by the regular block generation rate. As one regular reward means that a main block is generated, we can calculate the regular block generation rate by the regular reward. Since the number of transactions in a main block is different in each BTC blockchain, we assume that there is only one transaction in a main block, which is different from [14] and [15]. Thus, the maximum TPS is 1.

### 3.3.4 Hazard index

We propose *Hazard Index* (HI), namely, extra relative revenue per computing power, to quantify the severity of an attack. HI is defined as the ratio of extra relative revenue to honest-mining relative revenue. Let $RR_{HM}$ denote relative revenue if the attackers mine honestly, and $C_M$ denote the proportion of the computing power of the attackers. Attackers' honest-mining relative revenue is equal to the proportion of their computing power. Eq. (6) gives the formula for calculating HI.

$$HI = \frac{RR_M - RR_{HM}}{RR_{HM}} = \frac{RR_M - C_M}{C_M} \quad (6)$$

The range of HI is $[-1, +\infty)$. HI is 0 if miners behave honestly, indicating no harm to the system. HI is positive if the miner conducts attacks and gets extra profit, which damages system fairness and security. HI can be negative when the miner losses its deserved revenue. Negative HI does not mean benefit or no damage to the system. It indicates the probability that the attacker does not launch the attack. A larger HI represents a higher rate of reward, indicating higher severity of an attack. Although HI is still related to the computing power of attacker, it is negatively correlated with computing power.

The existing works [8][9][12]-[16] used RR to evaluate both attacker profit and attack severity. If $RR_M > RR_{HM}$, attacker launches attack to get extra profit and the attack is harmful to blockchain system. If $RR_M < RR_{HM}$, attacker does not launch attack. The difference between $RR_M$ and $RR_{HM}$ represents the loss of attacker. The reason that RR cannot be compared and HI can be compared directly is as follows.

- If a small pool and a large pool have the same RR by conducting attacks, the small pool gets more extra revenue than the large pool. HI of the small pool is larger than that of the large pool.
- If two pools get the same HI by conducting attacks, RR of a small pool is lower than that of a large pool, but the reward rates of two pools are the same.

RR of a large pool is naturally higher than a small pool. Thus, the increase in RR can be caused by the higher computing power or the attack itself. That is, RR cannot be used to compare the threat severity of different attack strategies unless the computing power of the attacker is fixed. On the contrary, HI reflects attack severity and can make comparisons directly.

## 4 EXPERIMENT RESULTS

This section first gives parameter settings in Section 4.1. Then we demonstrate the approximate accuracy of our model and formulas in Section 4.2. Section 4.3 illustrates the impact of stubborn mining attack in terms of attacker profitability, system TPS and hazard index. Experiment results of comparing stubborn mining attack with other attacks are presented in Section 4.4.

### 4.1 Configuration of System and Parameters

As explained in Section 3.1, the computing power of MP varies from 5% (0.05) to 45% (0.45). The recent 2016 main blocks in Bitcoin, Bitcoin Cash and Litecoin indicate that the fork probability of these BTC blockchains is around 0.6%~1.2% [23][26]. Although these blockchains generate blocks within minutes, there are also PoW blockchains producing blocks in seconds, such as PoW Ethereum and Bitcoin Fast. They have a higher fork probability of around 5% [14][15]. To investigate the blockchain system comprehensively, we set the fork probability $\theta$ to 0.5%, 1% (low fork probability), 5%, 10% (high fork probability), and 20% (super high fork probability). Other parameter settings are shown in TABLE 3. The numerical experiments are conducted in MAPLE [28].

TABLE 3
PARAMETER SETTINGS

| Notation | Setting | Notation | Setting |
|---|---|---|---|
| $\gamma_n$ | $1/n$ | $g_3^A$ | $\gamma_3^2$ |
| $g_2^A$ | $\gamma_2^2$ | $g_3^{AH}$ | $2\gamma_3(1-\gamma_3)$ |
| $g_2^{AH}$ | $2\gamma_2(1-\gamma_2)$ | $g_3^H$ | $(1-\gamma_3)^2$ |
| $g_2^H$ | $(1-\gamma_2)^2$ | $g_3^{H1}, g_3^{H2}$ | $(1-\gamma_3)^2/2$ |

### 4.2 Verification of Model and Formulas

We verify the proposed model and formulas in terms of relative revenue from two aspects: 1) compare with existing works; and 2) compare with simulation results.

Our literature review indicates that only the work in [16] (denoted as Yang model) is similar to ours. We reproduce their experiments and compare them with our work. Yang model only considered selfish mining attack in imperfect GHOST blockchain systems. Our model and formulas can also evaluate selfish mining attack and Yang model can be seen as a special case of our system. By setting $L$, $F$, $T = 0$, 0, 0 and setting other parameters of our model as same as Yang's, we can get Yang's results. The comparison of our results and Yang's results is given in Fig.3. "RRM" and "RRH" denote the relative revenue of MP and HPs, respectively. "-num" and "-Yang" denote the results of our model and Yang model, respectively. The scatter points fall almost exactly on the curve, indicating that our results are the same as Yang's, verifying our model's approximate accuracy from one aspect.

To verify our model and formulas in more general scenarios, we develop a simulator in C language and conduct a series of simulation experiments. In each round of the simulation, a blockchain is created with 10 million main chain blocks and we do 100 rounds for each group of parameters. Because of space limitation, we only give the results of the selfish mining attack and *LFT* strategy. Other stubborn strategies have similar results. Fig.3 and Fig.4 show the results, where "-sim" denotes simulation results. We observe that numerical and simulation results are very close under different fork probabilities. All these results indicate the approximate accuracy of our model and formulas, and then the correctness of our quantitative analysis approach is verified.

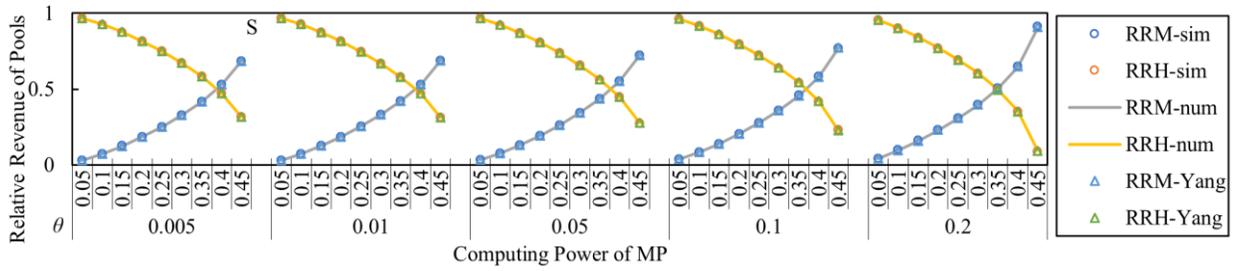

Fig.3 Verify our model and formulas under selfish mining

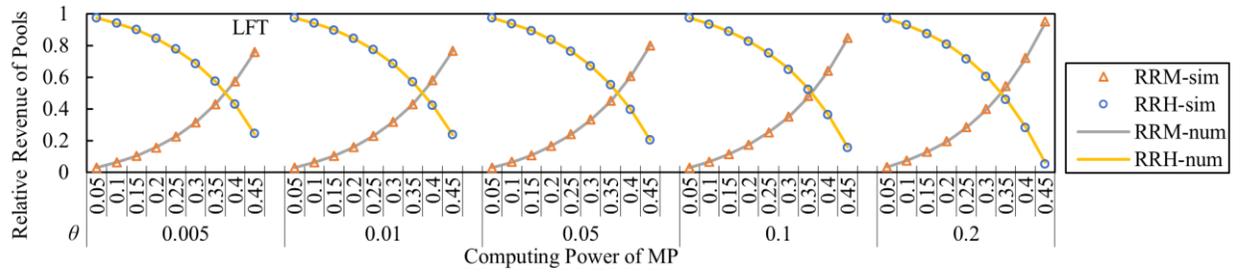

Fig.4 Verify our model and formulas under *LFT* strategy

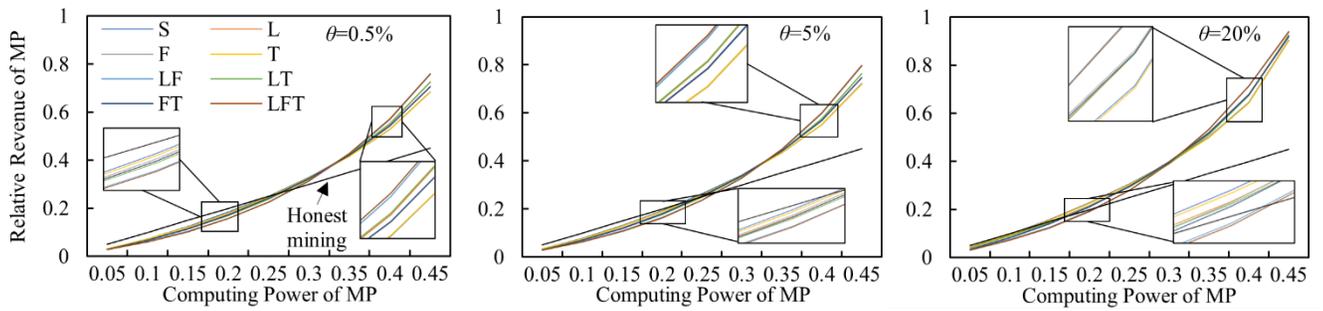

a) Fork probability is 0.5%    b) Fork probability is 5%    c) Fork probability is 20%

Fig.5 Relative revenue of MP under each stubborn strategy and selfish mining attack over computing power

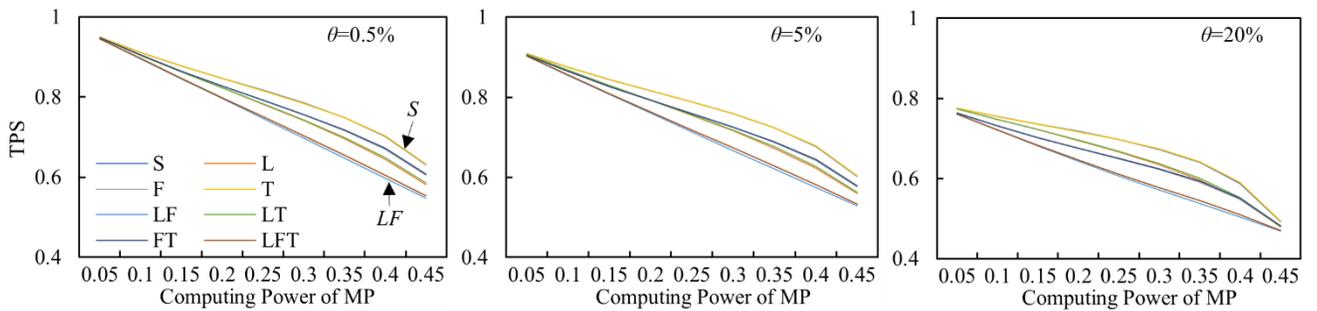

a) Fork probability is 0.5%    b) Fork probability is 5%    c) Fork probability is 20%

Fig.6 System TPS under each stubborn strategy and selfish mining attack over computing power of MP

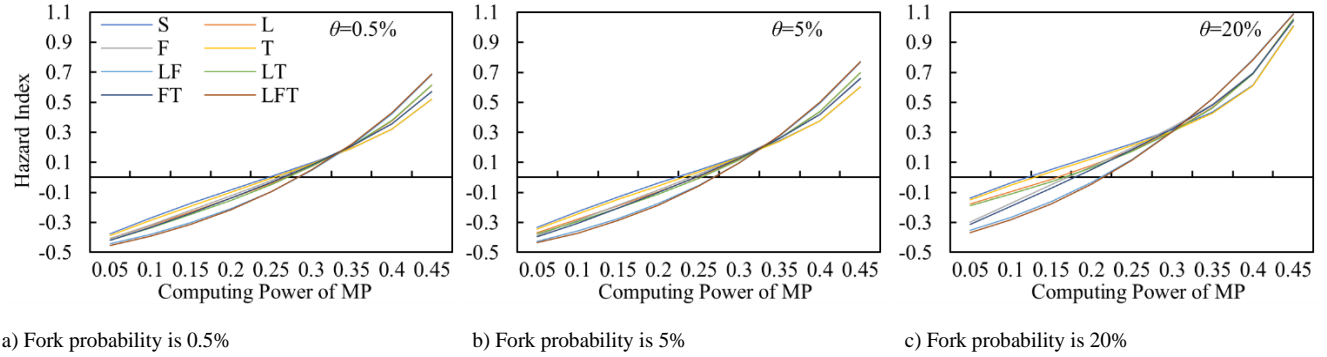

a) Fork probability is 0.5%    b) Fork probability is 5%    c) Fork probability is 20%

Fig.7 Hazard index of selfish and stubborn mining attack over computing power of MP

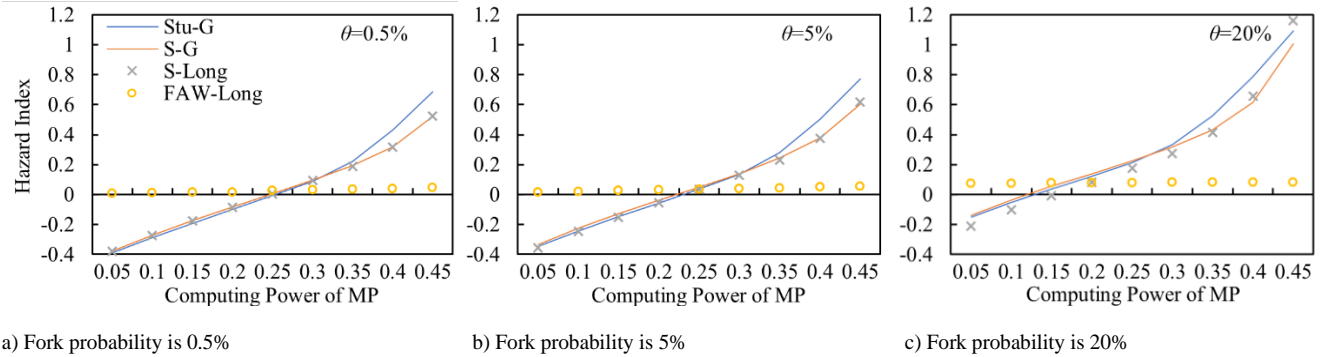

a) Fork probability is 0.5%    b) Fork probability is 5%    c) Fork probability is 20%

Fig.8 Comparison of hazard index among different attacks over computing power of MP

### 4.3 Evaluation of the Impact of Stubborn Mining Attack

This section first compares relative revenue of MP by varying fork probability. Then we evaluate the impact on system throughput and assess stubborn mining attack in terms of hazard index. We do experiments under a variety of fork probabilities mentioned in Section 4.1. Due to the space limitation, we choose three typical values of fork probability 0.5%, 5% and 20% to illustrate the impact of stubborn mining attack in low, high and super high fork probability, respectively.

#### 4.3.1 Relative revenue

Fig.5 shows the results of relative revenue of MP, where "S" denotes the results of selfish mining attack. We find that:

- High fork probability benefits attackers. Relative revenue of MP increases with the increasing fork probability no matter which strategy MP takes. This is demonstrated by all curves rising as the fork probability grows when Fig.5 (a)-(c) are compared horizontally. Adversary has the motivation to launch network attacks to increase fork probability.
- Selfish mining attack is more profitable than stubborn mining attack for small pools. This is demonstrated by the fact that the curve of selfish mining attack is higher than all stubborn strategies when the computing power of MP is less than around 30% (0.31 when $\theta=0.5\%$, 0.3 when $\theta=5\%$ and 0.28 when $\theta=20\%$) in Fig.5 (a)-(c). The reason is that selfish mining attack makes miners try to grasp the advantages in block leading while stubborn mining attack tends to make up for lagging behind. For example, when $\Delta=2$ and HPs find a block, MP publishes all two private blocks in selfish mining but only publishes one block in $L$ strategy. $L$ strategy hopes HPs to mine on private subtree, dispersing and wasting HPs' computing power. However, $L$ strategy increases the probability of losing advantage in block leading. HPs may generate a fork on same public subtree and then MP has to give up the second private block for falling behind. MP has a low probability of having two more blocks than public subtree when its computing power is small. Thus, stubborn mining attack is not good for small computing power attackers.
- Stubborn mining attack performs better than selfish mining attack for large pools. When MP has more than 30% computing power, the curve of selfish mining attack is the lowest in Fig.5 (a) and Fig.5 (b) and is the second lowest in Fig.5 (c). Pools with large computing power are more likely to generate valid blocks so they have a high possibility to catch up from lagging. Thus, stubborn mining strategies show their strengths in terms of relative revenue for the attackers.
- *T strategy (LT, FT, LFT) is more profitable than * strategy (L, F, LF) for attackers in most cases. This is demonstrated by the results that the curve of "*T" is very close to and slightly lower than that of "*" (namely, *LT* and *L*, *FT* and *F*, *LFT* and *LF*, and *T* and *S*) in Fig.5 (a)-(c). In other words, attackers should choose * strategy instead of *T strategy if they are unsure about the current fork probability and their computing power. However, *LFT* strategy does perform best when MP's computing power is 34%~45% if $\theta \leq 10\%$ in Fig.5 (a) and (b). If the attackers want to get the most profit in these scenarios, they still should adopt *LFT* strategy.

### 4.3.2 System throughput

Fig.6 gives the results of system TPS under each stubborn strategy over computing power of MP. Since the regular block generation rate and the block generation rate is normalized to 1 in Section 3.3, the maximum TPS of the system is 1. Fig.6 results show that:

- TPS decreases with the increasing computing power of MP and the higher fork probability. As shown in Fig.6 (a)-(c), all curves go downward with the increasing computing power of MP. The curves in Fig.6 (a) are higher than that in Fig.6 (b). This is because higher fork probability implies more stale blocks. Stubborn mining attack produces forks intentionally. Note that the block generation rate is stable. The number of main blocks decreases if there are more stale blocks within a same time period.
- TPS of *LF* and *LFT* are the smallest and TPS of selfish mining attack and *T* strategy are the largest in all fork probabilities no matter how much computing power MP holds. The reason is that stubborn strategies make more intentional forks than selfish mining attack, especially *L** and *F** strategies. In addition, the curves of *S*, *L*, *F*, *LF* are very close to and slightly higher than that of *T*, *LT*, *FT*, *LFT*, respectively (shown in Fig.6 (a)-(c)). *T strategy does less damage than * strategy in terms of TPS in most cases.
- TPS of *L* is higher than that of *F* for small MP and the opposite for large MP. This phenomenon is more obvious in Fig.6 (b) and (c). This is because Δ is more likely equal to 0 for small MP and *F* strategy works. For large MP, it is easier to produce blocks and has a heavier private subtree. Thus, the probability of Δ=2 is larger than that for small MP and *L* strategy works.

### 4.3.3 Hazard index

Metric hazard index (HI) reflects the attack damage per computing power caused to the blockchain system. It is related to the computing power of attackers and can tell an attacker whether to launch an attack: 1) Positive HI suggests an attacker to launch the attack and get extra revenue; and 2) negative HI means the loss of deserved revenue of the attacker and then rational attackers will not launch the attack. Note that negative HI does not mean benefit or no damage to the system. It indicates the probability that the attacker will not launch the attack. Fig.7 shows HI of stubborn mining and selfish mining attacks over computing power of MP. We observe that:

- Stubborn mining attack launched by large pools does more harm to the system. HI increases with the growth of the computing power of MP in Fig.7 (a)-(c). It also indicates that large pools gain more revenue by conducting stubborn mining attack.
- It is not a profitable choice for small pools to launch stubborn mining attack. This is demonstrated by the fact that HI of stubborn mining attack is negative when the computing power of MP is lower than around 25% in low (in Fig.7 (a)) and high fork probability (in Fig.7 (b)) and around 18% for super high fork probability (in Fig.7 (c)). The benefit threshold decreases with the increasing fork probability. Specifically, the threshold is 25%~30% when $\theta = 0.5\%$ and 23%~27% when $\theta = 5\%$. When $\theta = 20\%$, the threshold is 12% for *S* and *T*, 22% for *LF* and *LFT*, and 18%~20% for other strategies.
- The attackers in a blockchain with high fork probability tend to launch stubborn mining attack to get extra revenue. This is demonstrated by the results that HI increases with the increasing fork probability in Fig.7 (a)-(c). As shown in Fig.7 (c), MP with 45% computing power can double its relative revenue by stubborn mining attack when $\theta = 20\%$.
- Different from RR, HI can effectively assist attacker in finding out how profitable it is to conduct an attack. When RR of attacking is lower than that of honest mining, the existing works make difference to calculate the loss of MP. In Fig.5 (a), RR of MP adopting *L* strategy is 0.03 when the computing power of MP is 0.05, and RR is 0.115 when the computing power of MP is 0.15. MP with 0.05 computing power loses less (0.02) than the MP with 0.15 computing power (0.035). MP with 0.05 computing power may conduct the attack because the loss is little. However, in Fig.7 (a), HI of MP adopting *L* strategy is -0.41 for 0.05 computing power MP and is -0.23 for 0.15 computing power MP. MP with 0.05 computing power loses much more than MP with 0.15 computing power in reality and it does not conduct the attack.

### 4.4 Comparison of Stubborn Mining Attack and Other Attacks in Hazard Index

In an imperfect network scenario, there is no analytical model for stubborn mining attack under the longest-chain protocol and only selfish mining attack dynamics have been modeled under GHOST. Thus, we evaluate the damage severity of stubborn mining attack in GHOST by comparing with other blockchain attacks listed as follow:

1) Selfish mining attack under the longest-chain protocol. It is modeled in [14].
2) Selfish mining attack under GHOST. It can be modeled by both our work and the work in [16] and results are the same as shown in Section 4.2.
3) Fork after withholding (FAW) attack under the longest-chain protocol. It is modeled in [31]. Block withholding (BWH) attack [29] is a classic attack that is easy to conduct and hard to detect. Fork after with-holding (FAW) attack [30] is one of the most threatening BWH-based attacks, enabling attackers to always get extra relative revenue. Thus, we compare stubborn mining attack and FAW attack to investigate their damage.

We implement these three models and the related metric formulas. Fig.8 shows the results, where "S" denotes selfish mining, "Stu" denotes the optimal stubborn mining strategy (the strategy that maximizes HI), "-G" and "-Long" denote GHOST and the longest-chain respectively. We observe that:

- Stubborn mining attack is more harmful than selfish mining attack. In GHOST systems, HI of optimal stubborn strategy (denoted as opt-stubborn) is obviously higher than that of selfish mining attack when the computing power of MP is larger than 35% in Fig.8 (a)-(c). When the MP holds lower than 35% computing power, HI of opt-stubborn is slightly less than that of selfish mining attack. Moreover, the rewarding threshold of opt-stubborn and selfish mining attack is almost same. opt-stubborn outperforms selfish mining attack in terms of the severity to the blockchain.
- Small pools tend to launch FAW attack while large pools tend to launch stubborn mining attack. In the longest-chain system, HI of FAW attack is always positive and fluctuates little with the increase of MP's computing power. But the value is small (HI<0.05 when $\theta=0.5\%$ in Fig.8 (a) and HI < 0.085 when $\theta=20\%$ in Fig.8 (c)). Selfish mining attack causes much more damage when the computing power of MP is larger than 25%. HIs of selfish mining and opt-stubborn are negative if MP holds less than 25% computing power in Fig.8 (a) and (b).
- GHOST performs better than the longest-chain protocol in defending against selfish mining attack from large pools, but performs worse in resisting attack from small pools. The comparison results of GHOST and the longest-chain indicates that HIs of selfish mining are very close in low fork probability (0.5%) networks in Fig.8 (a). As shown in Fig.8 (b), in high fork probability, HI of GHOST selfish is higher than that of the longest-chain in scenarios where the computing power of MP is less than 35%. GHOST performs worse than the longest-chain in resisting small pool attacks. When MP has more than 40% computing power, HI of the longest-chain is greater than that of GHOST. This phenomenon still occurs in the situation of super high fork probability systems (shown in Fig.8 (c)).

## 5 CONCLUSION AND FUTURE WORK

This paper develops a Markov model to evaluate stubborn mining attack in imperfect networks. We propose a new metric called "Hazard Index" to quantitatively measure attacking damage severity. Metric formulas of relative revenue of the attacker, system throughput and Hazard Index are established. Based on these formulas, we make a quantitative study of stubborn mining attack in imperfect BTC blockchains, and we also compare it with selfish mining attack in GHOST, and with both selfish-mining and FAW attacks in the longest-chain scenario by hazard index.